\documentclass[12pt]{iopart}

\usepackage{graphicx}
\usepackage{color}
\usepackage{dcolumn}
\usepackage{bm}
\usepackage{amssymb}
\usepackage{mathrsfs}
\usepackage{braket}
\usepackage{multirow}
\usepackage{subfigure}
\usepackage{graphicx}
\usepackage[inline]{enumitem}
\begin{document}

\title{Conjugate two-dimensional electric potential maps}

\author{A. Bad\'{\i}a--Maj\'{o}s}
\ead{anabadia@unizar.es}
%\thanks{Corresponding author}
\address{Departamento de F\'{\i}sica de la Materia
Condensada and Instituto de Ciencia de Materiales de Arag\'{o}n (ICMA), Universidad de Zaragoza--CSIC, 
Mar\'{\i}a de Luna 3, E-50018 Zaragoza, Spain}
\author{E. de Lorenzo Poza}
\address{Facultad de Ciencias, Universidad de Zaragoza, Pedro Cerbuna 12, E-50009 Zaragoza, Spain}

%\date{\today}

\vspace{10pt}
\begin{indented}
\item[]July 29, 2018
\end{indented}

\begin{abstract}

{Two dimensional electric potential maps based on voltage detection in conducting paper are common practice in many physics courses in college. Most frequently, students work on ``capacitor-like'' geometries with current flowing between two opposite electrodes. A ``topographical'' investigation across the embedding medium (map of equipotential curves) allows to reassure a number of physical properties.

This paper focuses on some less common configurations that bear pedagogical interest. We analyze ``open-geometries'' with electrodes in the form of long strips with slits. They provide a natural groundwork to bring the student to complex variable methods. Aided by this, we show that shaping the conducting paper board one may analyze finite size effects, as well as some meaningful discontinuities in the measured potential.

The concept of conjugate electric potentials is exploited. Equipotentials and electric field lines acquire interchangeable roles and may be obtained in complementary ``dual'' experiments. A feasible theoretical analysis based on introductory complex variables and standardized numerics gives a remarkable quantification of the experimental results.}

\end{abstract}

\vspace{1cm}
\submitto{European Journal of Physics}
\maketitle

\section{Introduction}

{The electromagnetic theory education is full of challenges for the instructor. In particular, the comprehension of such prominent concepts as the nature and {properties} of scalar and vector field quantities is tough. Therefore, accompanying pedagogical resources do always play a relevant role.}

Two-dimensional field mapping has been spotted as a readily available experiment that provides a very helpful tool in physics courses \cite{ref:young1,ref:young2,ref:awk,ref:watta}. Thus, merely relying on silver painted electrodes over a sheet of weakly conducting (carbon impregnated) paper, a power supply (even just a battery!), and a voltmeter, one may introduce a number of fundamental ideas on the electric field $\bf E$ and potential $\Phi$. Recall that $\Phi$ is straightforwardly measured by scanning the paper with a single voltage probe, and that the lines of $\bf E$ may be plotted by holding one tap of the probe on one point of the paper and rotating the other one around until a maximum potential difference is found \cite{ref:pasco}.
Based on such experiments, it may be straightforwardly visualized that: \begin{enumerate*}
\item[\bf (a)] the potential is constant over the ``ideal'' conducting material (typically silver), 
\item[\bf (b)] the equipotential lines change smoothly, nearly reproducing the shape of the electrodes at small distances,
\item[\bf (c)] the equipotentials become closer and closer as one approaches the conductors Additionally, one may
\item[\bf (d)] assess the direction of the electric field vector at different points on the board (perpendicular to displacements $d{\bm \ell}$ along the equipotentials), and
\item[\bf (e)] check out that the electric field is perpendicular to the boundary of conductors.
\end{enumerate*}

Also, the student may be asked to quantify some properties. Thus, one may invoke the relation between the electric field and potential ($d\Phi = -{\bf E}\cdot d{\bm \ell}$) and:
\begin{enumerate*}
\item[\bf (f\,)] obtain the modulus of the electric field from the ratio between the voltage increment and the distance between closeby equipotentials,
\item[\bf (g)] estimate the net charge on the surface of the conductors as an application of Gauss' law (by addition of the electric field values, times the surface elements along a closed equipotential contour around the electrodes, i.e.: $Q\approx \epsilon_0 \sum_i E_i \Delta S_i$).
\end{enumerate*}

All the above can be well exploited in the framework of Introductory Physics courses. Noteworthily, this experiment brings the student far beyond the  ``academic'' situation of ideal point charges and Coulomb's law. The electric potential is revealed as a highly important concept, in particular as a link among the underlying ``invisible'' agents, i.e.: electric charges, and measurable quantities in the lab. 

Further added value of potential maps may be obtained by taking advantage of more specialized techniques as those in intermediate electromagnetism courses and be implemented in upper division laboratories \cite{ref:young1,ref:young2,ref:awk}. Thus, unless for the case of circular concentric electrodes, that trivially gives way to circular equipotentials, other typical setups must be treated either by numerical solution of Laplace's equation, image methods, complex variables\dots\; at least if an exact solution is wanted!
Eventually, relying on such methods, the student may compare the theoretical and observed equipotentials, check the calculated capacity of the system {\em vs} the measured value, {\em etc}.

Still, as we show along this paper, the use of potential maps allows new opportunities to upgrade the knowledge on concepts and methods for electromagnetic education. On the one side, we will show that the introduction of ``open electrode geometries'' in the shape of long strips with slits provides a groundwork either to bring the student to complex variable methods in electrostatics or to strengthen their knowledge. This technique was a major and elegant tool in the analysis of 2D potential problems decades ago (see Ref.\,\cite{ref:smythe} for a thorough treatment). In the last years, although present in the syllabus of BSc in Physics, complex variable methods are less touched beyond the mathematical subjects themselves. Nevertheless, some recent works \cite{ref:young2,ref:awk,ref:jacksonAJP} put forward their revival. Along such line, in this article, we will expose the idea of {\em reciprocity} \cite{ref:awk,ref:conjugates,ref:beth}, i.e.: electric field lines and equipotentials may be represented by  ``dual'' harmonic conjugate functions of a complex potential. This mathematical property will reveal a striking physical effect. One may design complementary laboratory experiments for which equipotentials and ${\bf E}-$lines interchange roles.
Also, conveniently interpreted, this feature will allow us to extend our analysis to \begin{enumerate*}
\item[\bf (h)] the quantification of finite size effects, and \item[\bf (i)] to introduce in practice such relevant concepts as continuity/discontinuity and electrical coupling/uncoupling.
\end{enumerate*}

The plan of the paper is as follows. First, in order to make the work self-contained, we will describe the physical problem and its conventional analysis in the realm of a standard calculus-based course of Electricity (Sec.\,\ref{sec_IIB}). Then, in Sec.\,\ref{sec_IIC}, we upgrade the mathematical treatment. It is devoted to introduce some basic concepts on complex variable techniques. A brief description of some accessible numerical methods and their application to analyze finite size effects is also included (Secs.\,\ref{sec_IID}, \ref{sec_IIE}). In order to make the presentation clearer, methodologies are illustrated through the application to specific configurations of interest. Sec.\,\ref{sec_III} presents our own experimental measurements in the standard setup that is used in our Physics lab. The analysis of results in the light of the previous theoretical concepts, together with some discussion about possible extensions closes the article (Sec.\,\ref{sec_IV}).

\section{Theoretical background}
\label{sec_II}
\begin{figure*}[t]
\centering
{\includegraphics[width=.45\textwidth]{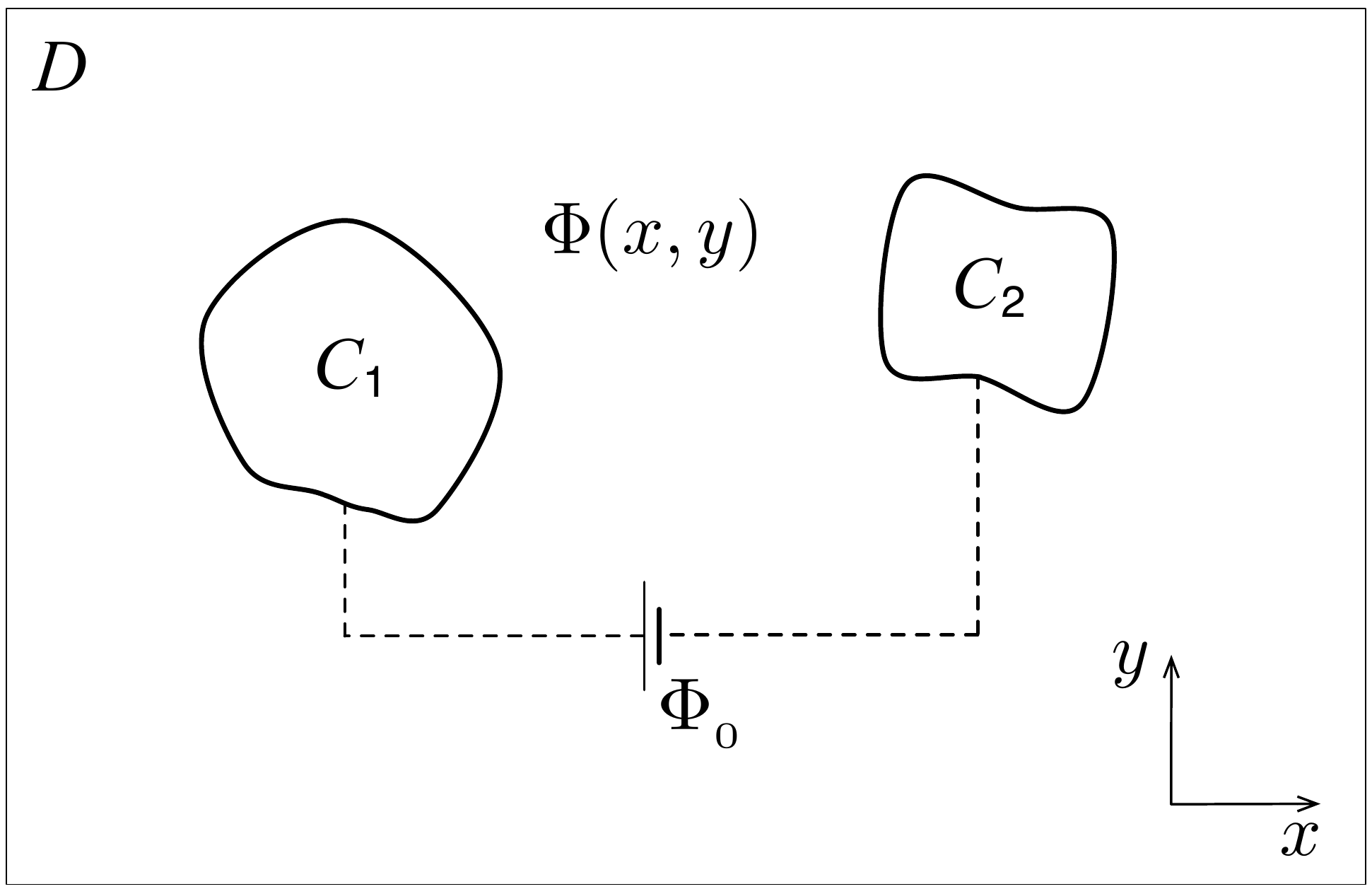}}
\caption[1]{\label{Fig_1}
Two perfect conductors $C_1 , C_2$ are held at a potential difference $\Phi_0$. An electric field is established in the embedding medium $D$, with associated potential function $\Phi (x,y)$.}
\end{figure*}

As sketched in Fig.\,\ref{Fig_1}, standard electric potential 2D mapping relies on finding the {\em loci} of constant potential lines around a given configuration of silver paint terminals over a conducting sheet. Thus, one records families of points fulfilling
\begin{equation}
\Phi(x_{ij},y_{ij})= k_j \,\Phi_0
\end{equation}
with $\Phi_0$ the potential difference between terminals, $k_j$ a given fraction of unity, and $i$ running along the set of points in a given equipotential (labelled with $j$). This is a sampling of the electric potential $\Phi$, that will be represented by the two-variable function $\Phi (x,y)$ within the region of interest in the $XY-$plane (our experimental board).

In the forthcoming paragraphs we review some basic mathematical background. Ahead, let us call to mind that usually $\Phi (x,y)$ is said to well represent either our actual conduction problem in the lab or its electrostatic analog. However, this statement has to be dealt with care if finite size effects are relevant (see supplementary material). Such fact will be analyzed and exploited in our suggested experiments (secs.\,\ref{sec_IIE} and \ref{sec_IIIB}). Equivalence will not be assumed by default.
%
%%%%%%%%%%%%%%%%%%%%%%%%%%%%%%%%%%%%%%%%%%%%%%%%%%%%%%%%%%%%%%%%%%%%%%%%%%%%%%%
%% FIGURE 1 
%%%%%%%%%%%%%%%%%%%%%%%%%%%%%%%%%%%%%%%%%%%%%%%%%%%%%%%%%%%%%%%%%%%%%%%%%%%%%%%
%%%%%%%%%%%%%%%%%%%%%%%%%%%%%%%%%%%%%%%%%%%%%%%%%%%%%%%%%%%%%%%%%%%%%%%%%%%%%%%%

\subsection{Geometrical interpretation: the conjugate potentials ($\Phi ,\zeta$)}
\label{sec_IIB}

Let us start by bringing back the differential relation between the electric field and potential
\begin{equation}
\displaystyle{{\bf E}=(E_x, E_y)= -\left(\frac{\partial \Phi}{\partial x}, \frac{\partial \Phi}{\partial y}\right)\equiv -{\bf grad}\,\Phi} \, ,
\end{equation}
which {relates} to the following expression for directional increments of the potential (along an arbitrary displacement $d{\bm\ell}$)
\begin{equation}
d\Phi ={\bf grad}\,\Phi \cdot d{\bm\ell}= -{\bf E}\cdot d{\bm\ell}=-E_x dx -E_y dy \, .
\end{equation}
Here, $d\Phi$ {is} the notation for the variation of the potential between neighbouring points, i.e.: $\Phi(x+dx,y+dy)-\Phi(x,y)$.

This leads to a useful geometrical interpretation. We start by recalling that equipotentials are defined by
\begin{equation}
\Phi= constant \Rightarrow d\Phi =0 \, ,
\end{equation}
which obviously means that ${\bf E}\perp d\bm{\ell}$ if displacements take place within a given contour level of $\Phi$.

Here, we want to notice that in 2D problems, one may complement $\Phi(x,y)$ with another relevant scalar field, $\zeta(x,y)$ in what follows. Conversely, it will be defined by the condition ${\bf E}\parallel d\bm{\ell}'$, with the displacement $ d\bm{\ell}'$ along the contour lines of the function $\zeta(x,y)$. Mathematically, this  means that ${\bf E}$ is perpendicular to the gradient of $\zeta$, which leads to
\begin{equation}
\displaystyle{{\bf E}\cdot{\bf grad}\,\zeta=0}
%\nonumber
%\\
\;\Rightarrow\;
%\nonumber
%\\
\displaystyle{{\bf grad}\,\Phi\cdot{\bf grad}\,\zeta=0}  \, .
\nonumber
\end{equation}
This condition establishes a relation (safe a sign) between partial derivatives. For convenience, we will use it in the form

\begin{equation}
\label{eq:C_R}
\frac{\partial \Phi}{\partial x}= -\frac{\partial \zeta}{\partial y}\quad ; \quad \frac{\partial \Phi}{\partial y}= \frac{\partial \zeta}{\partial x} \, .
\end{equation}

Below, we will show that the introduction of complex variables offers an advantageous unifying picture of these properties.
The function $\zeta (x,y)$ will be named after {\em flux} function, as its contour lines indicate the direction of {\em electric flow}.
As clarified later, one may refer to $\Phi$ and $\zeta$ as {\em conjugate} potentials. Notice that the Eq.(\ref{eq:C_R}) straightforwardly implies that, {under very general conditions, both $\Phi$ and $\zeta$ are harmonic functions ($\nabla^2 \Phi =0\, ,\, \nabla^2 \zeta =0$)}.

\subsection{Complex representation equations}
\label{sec_IIC}

Eqs.(\ref{eq:C_R}) may be recognized as the celebrated Cauchy-Riemann conditions \cite{ref:churchill} that are satisfied by harmonic conjugate functions. We proceed by considering the unique (safe constants) analytic function whose real and imaginary parts are $\zeta$ and $ \Phi$. {It} will be named after {\em complex potential}
\begin{equation}
\psi (z) = \zeta (x,y)+ i \Phi (x,y)
\end{equation}
over the complex plane ($z = x + i y$) \cite{ref: multivalued}. Here, we stress that $(x,y)$ represents a point on our experimental board. As in many other instances, the use of complex variables must be understood as a suitable representation to conveniently deal with the real valued physical variables.

Next, the evaluation of $\Phi (x,y)$ [equivalently of $\zeta (x,y)$] around the conductors may be done by taking advantage of some important mathematical properties. Firstly, being harmonic functions, they are uniquely determined by the (Dirichlet/Neumann) conditions along the boundaries of the region of interest (conducting paper). 
Below, we illustrate the use of such property in some specific arrangements. They will be formed by aligned electrodes with small slits in between. From the physical point of view, we will assume that length scales are such that, in the region of interest, all distances may be neglected as compared to the length of the electrodes (i.e.: thickness, width of the slits,...). Mathematically, we will treat the electrodes as lines and the slits as points.

To start with, and so as to gain familiarity with the complex representation, we put forward a couple of examples for which the solution of Laplace's equation is easily found. Further application is shown as supplementary material.

\subsubsection*{Example 1: long strip with single slit}

\begin{figure*}[t]
\centering
\subfigure{
\includegraphics[scale=0.45]{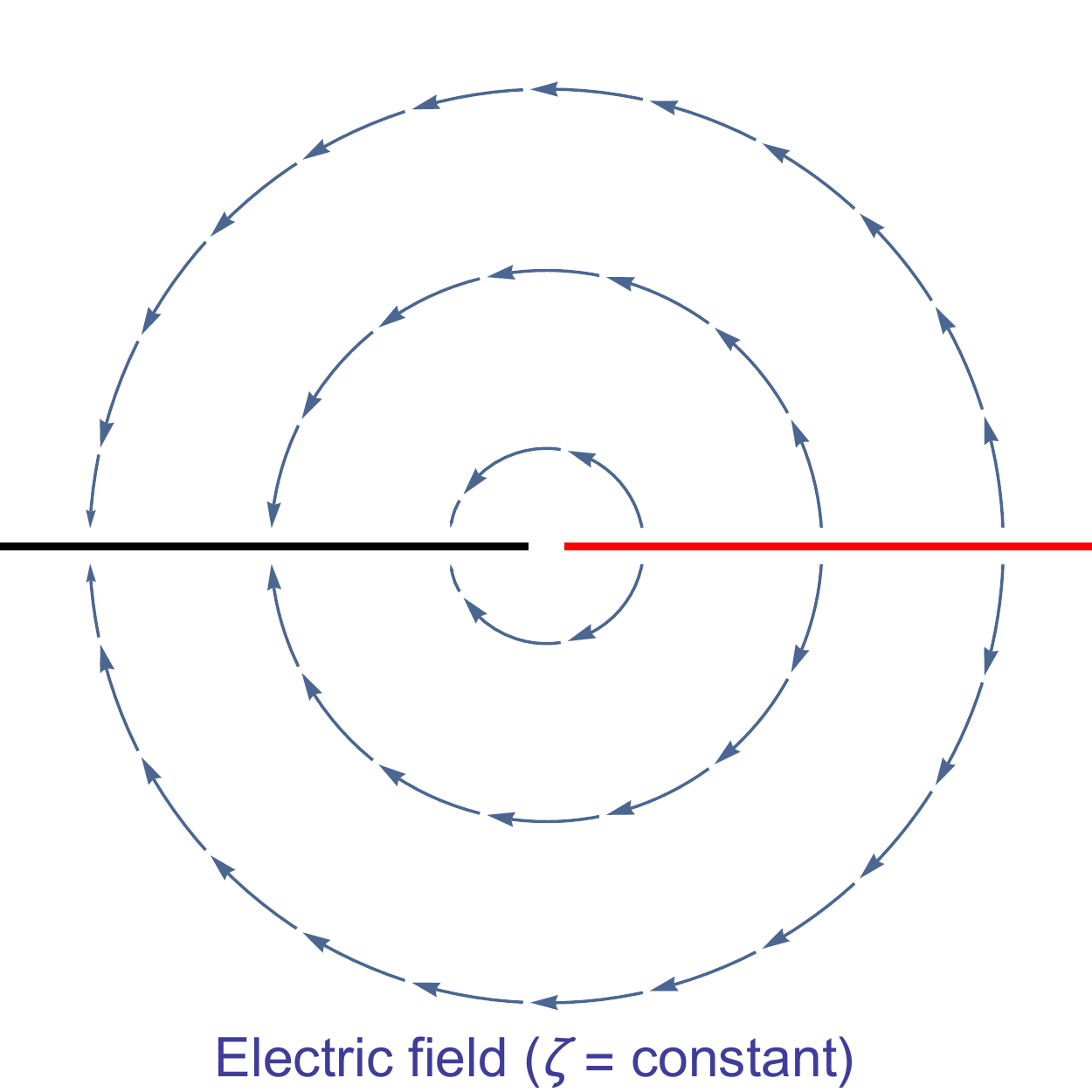}}
%\noindent\rule{8cm}{0.4pt}
\qquad\subfigure{
\includegraphics[scale=0.45]{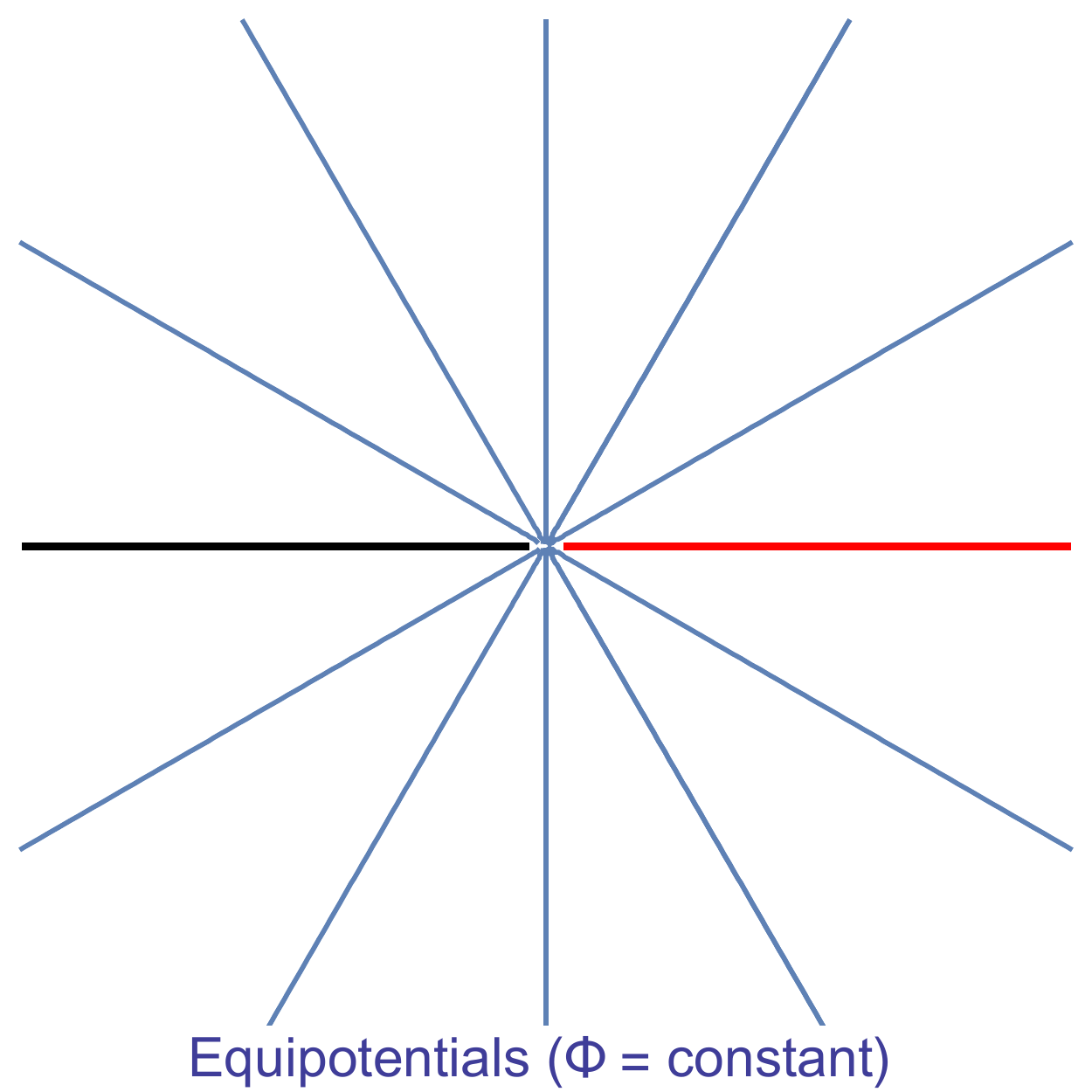}}
\caption{\label{Fig_2}(Color online)
 Electric field lines (left pane) and equipotentials (right pane) corresponding to the ``monopole'' situation generated by the single slit configuration, evaluated from the complex potential $\psi_m$ (see Eq.(\ref{eq:complex_mono})).}
\end{figure*}

Let us assume that our system may be approximated by two aligned electrodes in the form of an infinitely (in practice ``very'') long strip with a tiny central slit (Fig.\,\ref{Fig_2}). 
One of the electrodes is held at potential $0$ and the other one at the value $\Phi_0$.

First of all, the reader may straightforwardly check that the function  
\begin{eqnarray}
{\displaystyle \Phi(x,y)=\frac{\Phi_0}{\pi}\theta (x,y)=\frac{\Phi_0}{\pi}{\rm tan}^{-1}\left(\frac{y}{x}\right)}
\end{eqnarray}
with $\theta$ the polar angle, satisfies the Laplace equation (a trivial result using polar coordinates), and fulfills the boundary conditions for the upper half-plane $(y\geq 0)$, i.e.:
\begin{eqnarray}
{\displaystyle \Phi(r,\theta=0)=0\quad ,\quad \Phi(r,\theta=\pi)=\Phi_0}
\end{eqnarray}
independently of distance.

The solution for $y<0$ follows immediately by symmetry (or may be derived independently).

On the other hand, recall that in the complex plane $\theta (x,y)$ may be interpreted as the imaginary part of the natural logarithm \cite{ref:churchill}, i.e.
\begin{equation}
{\rm log}\,(z)={\rm log}\,|z| + i\,{\rm arg}\,(z) \equiv {\rm log}\,r + i\,\theta \,  .
\end{equation}
%
%%%%%%%%%%%%%%%%%%%%%%%%%%%%%%%%%%%%%%%%%%%%%%%%%%%%%%%%%%%%%%%%%%%%%%%%%%%%%%%
%% FIGURE 2 
%%%%%%%%%%%%%%%%%%%%%%%%%%%%%%%%%%%%%%%%%%%%%%%%%%%%%%%%%%%%%%%%%%%%%%%%%%%%%%%
%%%%%%%%%%%%%%%%%%%%%%%%%%%%%%%%%%%%%%%%%%%%%%%%%%%%%%%%%%%%%%%%%%%%%%%%%%%%%%%%
%
This is an analytic function for $y > 0$, and thus, the {\em flux} function will be given by its real part. A complex potential $\psi_m $ ($m$ standing for ``monopole'') may thus be defined
\begin{eqnarray}
\label{eq:complex_mono}
\fl
\displaystyle{\psi_m (z)=\frac{{\Phi}_0}{\pi}\,{\rm log}(z)}
\nonumber
\\
\Downarrow
\\
\fl
\displaystyle{{\zeta_m}={\rm Re}(\psi_m)=\frac{{\Phi}_0}{\pi}{\rm log}\,\sqrt{x^2+y^2}}
\quad ;\quad
\displaystyle{{\Phi_m}={\rm Im}(\psi_m)=\frac{{\Phi}_0}{\pi}\,{\rm tan}^{-1}\left(\frac{y}{x}\right)}
\nonumber
\end{eqnarray}

Fig.\,\ref{Fig_2} displays the physical quantities ${\bf E},\Phi$ derived from the above analysis for the single slit. They are respectively obtained as the contour lines of the functions $\zeta_m (x,y)$ and $\Phi_m (x,y)$.

Apparently, we obtain a structure that is ``reciprocal'' to  the most common problem of the potential around the tiny central conductor in a cylindrical capacitor geometry. Such system is straightforwardly solved by application of Gauss' law. The complex representation of its solution is 
\begin{equation}
{\displaystyle\tilde{\psi}_m(z)=\frac{\Phi_0}{{\rm log}(r_{_0})}{\rm log} (z)\equiv \tilde{\Phi}(x,y)+i\tilde{\zeta}(x,y)}
\end{equation}
with $\Phi_0$ the reference potential at the distance $r_{_0}$.
Notice that the roles of the functions giving the potential and electric flux are interchanged with our problem's. In this case, equipotentials are given by ${\tilde{\Phi}(x,y) \propto {\rm log}(r)= constant}$ whereas the electric flux lines correspond to ${\tilde{\zeta}(x,y)\propto\theta=constant}$.  In other words, our single slit system has got a dual system with the correspondence
\begin{equation}
\displaystyle{\tilde{\psi}_m(z)\,{=}\,i\stackrel{\rule{2.5em}{1pt}}{\psi_m (z)}}
\end{equation}

We want to stress that, though for a trivial system, the above result is somehow challenging. Just by ``painting'' the silver electrodes along two specific lines of the conventional monopole electric field, the reciprocal system maps the whole picture of ${\bf E}-$lines and equipotentials with interchanged roles. The question arises: does this property of the ``trivial'' monopole system apply to other distributions of interest in some manner?  Along this line, next, we upgrade the result by showing that one may also generate dipole field flux lines by a ``reciprocal'' electric potential mapping system.

\subsubsection*{Example 2: long strip with two slits}

\begin{figure*}[b]
\centering
\subfigure{
\includegraphics[scale=0.45]{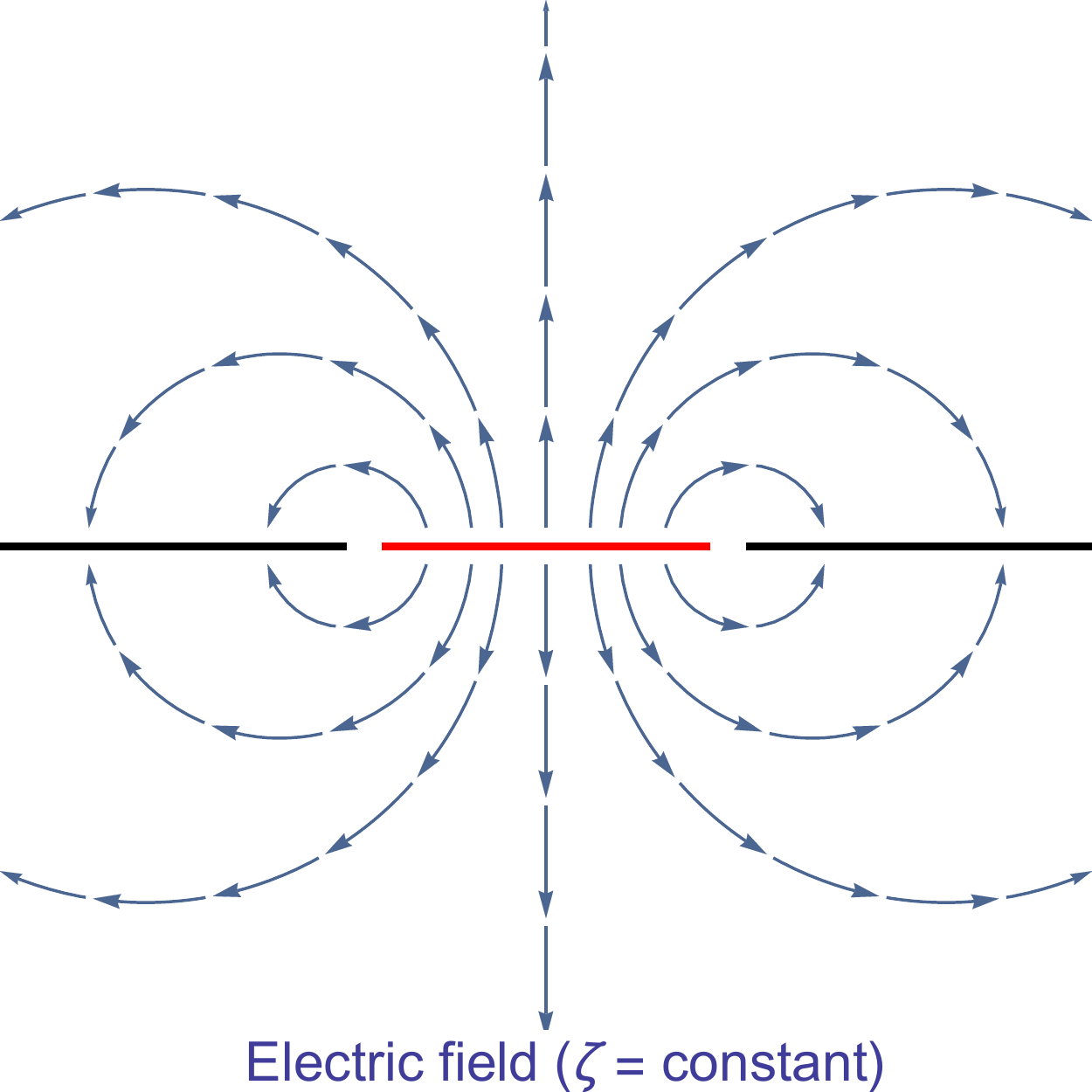}}
%\noindent\rule{8cm}{0.4pt}
\qquad\subfigure{
\includegraphics[scale=0.45]{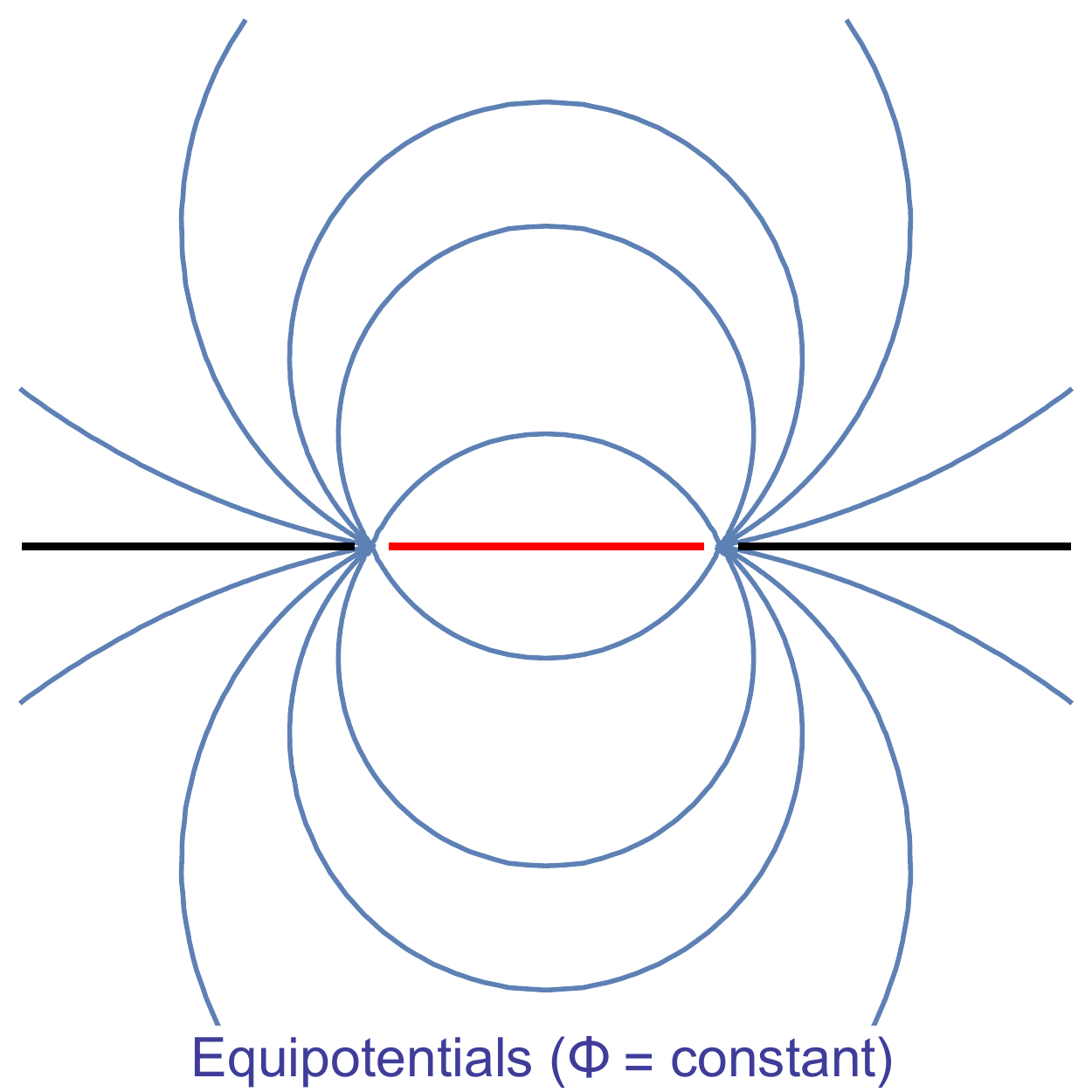}}
\caption{\label{Fig_3}(Color online)
 Electric field lines (left pane) and equipotentials (right pane) corresponding to the ``dipole'' situation generated by the double slit configuration, obtained from the complex potential $\psi_d$.}
\end{figure*}

As shown in Fig.\,\ref{Fig_3} our next example consists of a long strip interrupted by two slits. In certain units they may be located at $z=1$ and $z=-1$. Now, a combination of the functions that measure the polar angle with reference to each slit (say $\theta_1 , \theta_2$) will give the expression for the electrostatic potential on the upper half-plane. If, as indicated, one assumes the voltage connections to the conductors such that
\begin{eqnarray}
\Phi (x,0) = \left\{
\begin{array}{ll}
0\; &,\;  x<-1
\\
\Phi_0\; &,\;  -1<x<1
\\
0\; &,\;  x>1
\end{array}
\right. \, ,
\end{eqnarray}
also, a superposition of logarithms solves the problem. 
%
%%%%%%%%%%%%%%%%%%%%%%%%%%%%%%%%%%%%%%%%%%%%%%%%%%%%%%%%%%%%%%%%%%%%%%%%%%%%%%%
%% FIGURE 3 
%%%%%%%%%%%%%%%%%%%%%%%%%%%%%%%%%%%%%%%%%%%%%%%%%%%%%%%%%%%%%%%%%%%%%%%%%%%%%%%
%%%%%%%%%%%%%%%%%%%%%%%%%%%%%%%%%%%%%%%%%%%%%%%%%%%%%%%%%%%%%%%%%%%%%%%%%%%%%%%%
%
In this case, the appropriate complex potential is
\begin{eqnarray}
\label{eq:complex_dipo}
\fl
\displaystyle{\psi_d=\frac{{\Phi}_0}{\pi}\, \left[{\rm log}(z-1)-{\rm log}(z+1)\right]
=\frac{{\Phi}_0}{\pi}\, {\rm log}\frac{z-1}{z+1}}
\nonumber
\\
\Downarrow\nonumber
\\
\fl\displaystyle{{\zeta_d}=\frac{{\Phi}_0}{\pi}{\rm log}\,\sqrt{\frac{(x-1)^2+y^2}{(x+1)^2+y^2}}}
\quad ; \quad
\displaystyle{{\Phi_d}=\frac{{\Phi}_0}{\pi}{\rm tan}^{-1}\left(\frac{2y}{x^2+y^2-1}\right)}
%\nonumber
\end{eqnarray}
Notice that the function $\psi_d$ is analytic for $y > 0$. On the other hand, it is a simple exercise to show that (as wanted) ${\Phi_d}$ goes to $0, \Phi_0 ,0$ upon the conducting segments over the real axis. For this purpose, just notice that geometrically, ${\rm arg}(z\pm 1)$ may be visualized as the polar angle relative to the origin displaced at $z=\mp 1$ respectively.

Straightforward algebra shows that the equipotentials and electric flux lines are families of circles, respectively centered at the $y-$axis and $x-$axis. Thus, the electric field lines are given by 
\begin{equation}
\label{eq:circle_elec}
(x+b)^2+y^2=b^2-1
\end{equation}
with $b^2 \ge 1$ a constant, while equipotentials are given by 
\begin{equation}
\label{eq:circle_pot}
x^2 +(y-c)^2=1+c^2
\end{equation}
with $c$ a constant, that depends on the actual potential at a given contour. In fact, one may show that $c=1/{\rm tan}(\pi\Phi / \Phi_0)$.

As the reader might expect, the situation considered in this example is nothing but the ``conjugate'' of the electrostatic response for a 2D system consisting of two small conductors, with opposite charges, one at $z=1$ and the other at $z=-1$. Reciprocity is valid again, i.e.: 
$\tilde{\psi}_d(z)\,{=}\,i\stackrel{\rule{2.5em}{1pt}}{\psi_d (z)}$.

Higher order arrangements (i.e.: quadrupole-like) are suggested in the supplementary material.

\subsection{Numerical solution of Laplace's equation}
\label{sec_IID}

We call the readers' attention that the initiatory application of complex variable techniques brought in the previous section restricts to idealized domains, i.e.: the electrodes are semi-infinite lines and the conducting medium extends to infinity. Nevertheless, as we shall see, finite size effects in real systems may be very noticeable. In this section, this topic will be addressed through the numerical solution of Laplace's equation. A readily applicable method for obtaining $\Phi (x,y)$ based on widespread software will be described below \cite{ref:mathematica,ref:code}.

{\sc Mathematica}'s built-in functions allow us to easily define the differential equation, the domain, and the boundary conditions of the problem of which we want to find a numerical solution. In that sequence, we may define the differential equation with the function \texttt{Laplacian}, the region of interest may be expressed as a combination of basic regions (\texttt{Rectangle}, \texttt{Disk},...) through basic set operations (\texttt{RegionUnion}, \texttt{RegionDifference}, \texttt{RegionIntersection},...) and the boundary conditions, corresponding to the potential at the conductors introduced by \texttt{DirichletCondition}. It is also possible to explicitly state a perpendicularity condition for the potential [$\partial_{n}\Phi = F(x,y)$] with \texttt{NeumannValue}. It must be mentioned that the solver assumes null Neumann conditions ($\partial_{n}\Phi =0$) on the boundary of the domain unless otherwise indicated. In our case this fits the physical condition that electrons must flow along the boundary of the medium (carbon paper) and not across. Then, perpendicularity of equipotentials and electric flow leads to $\partial_{n}\Phi =0$.

%%%%%%%%%%%%%%%%%%%%%%%%%%%%%%%%%%%%%%%%%%%%%%%%%%%%%%%%%%%%%%%%%%%%%%%%%%%%%%%
%% FIGURE 4 
%%%%%%%%%%%%%%%%%%%%%%%%%%%%%%%%%%%%%%%%%%%%%%%%%%%%%%%%%%%%%%%%%%%%%%%%%%%%%%%
\begin{figure*}[t]
\centering
\subfigure{
\includegraphics[scale=0.3]{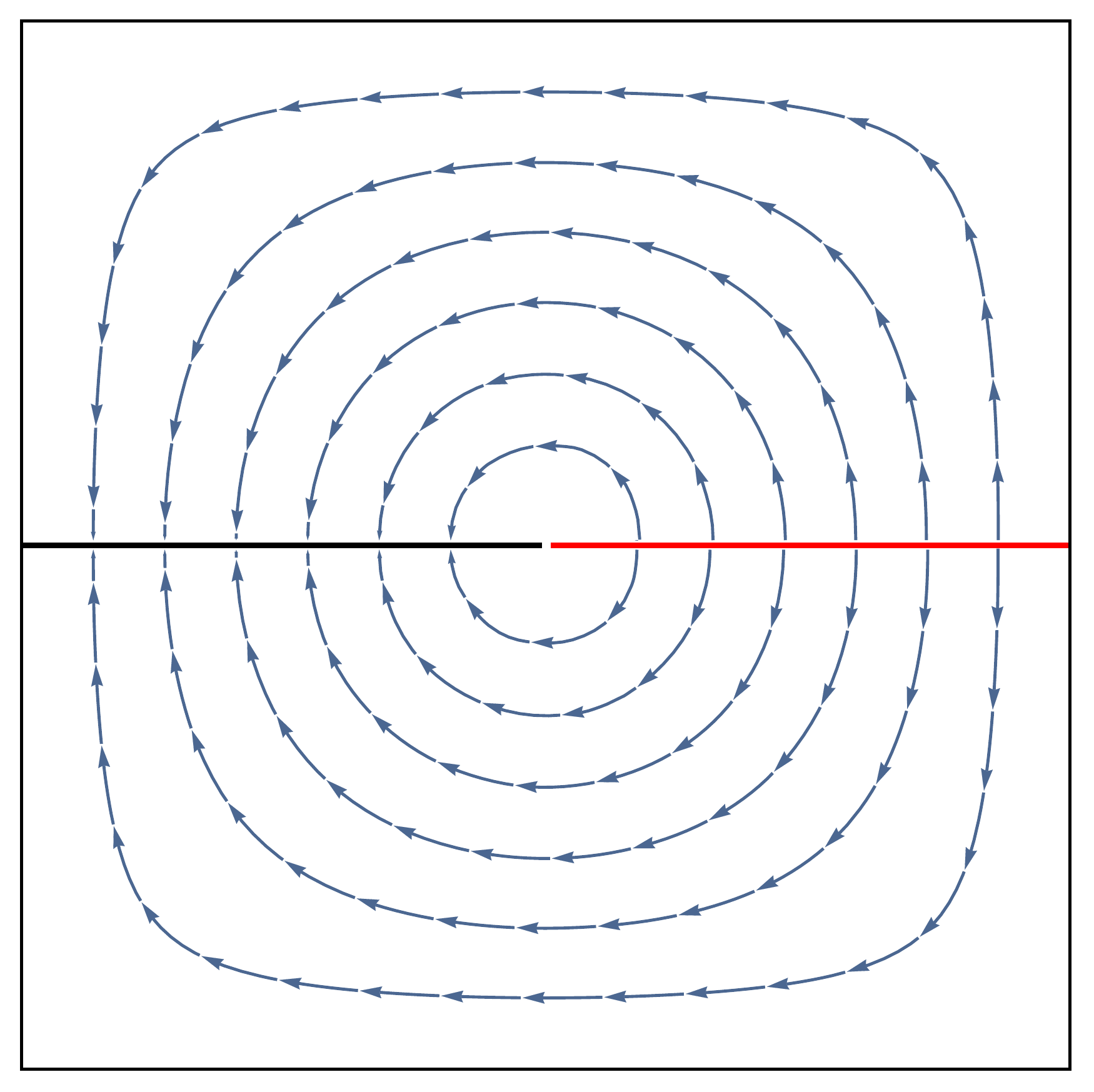}}
%\noindent\rule{8cm}{0.4pt}
\qquad\subfigure{
\raisebox{0.5ex}{\includegraphics[scale=0.67]{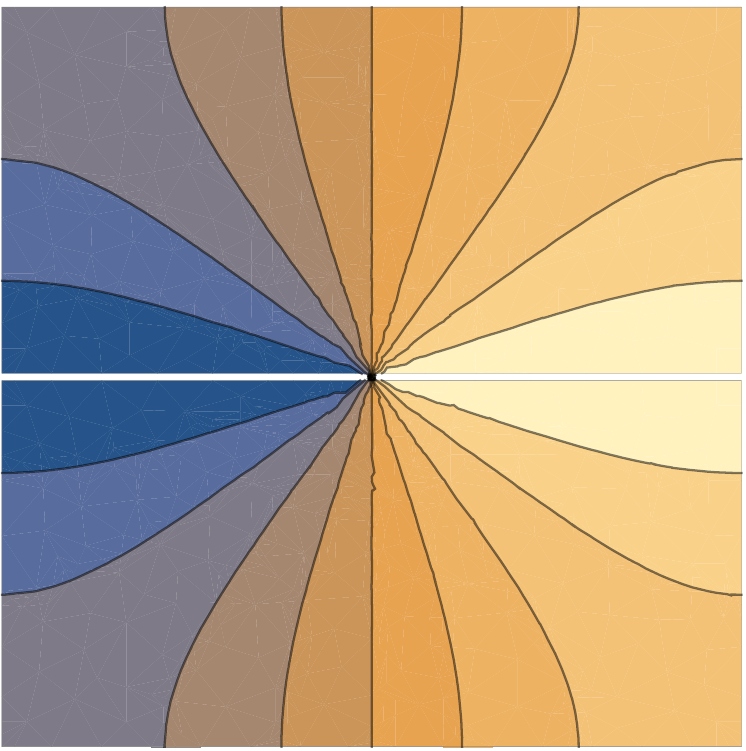}}}
\subfigure{
\includegraphics[scale=0.3]{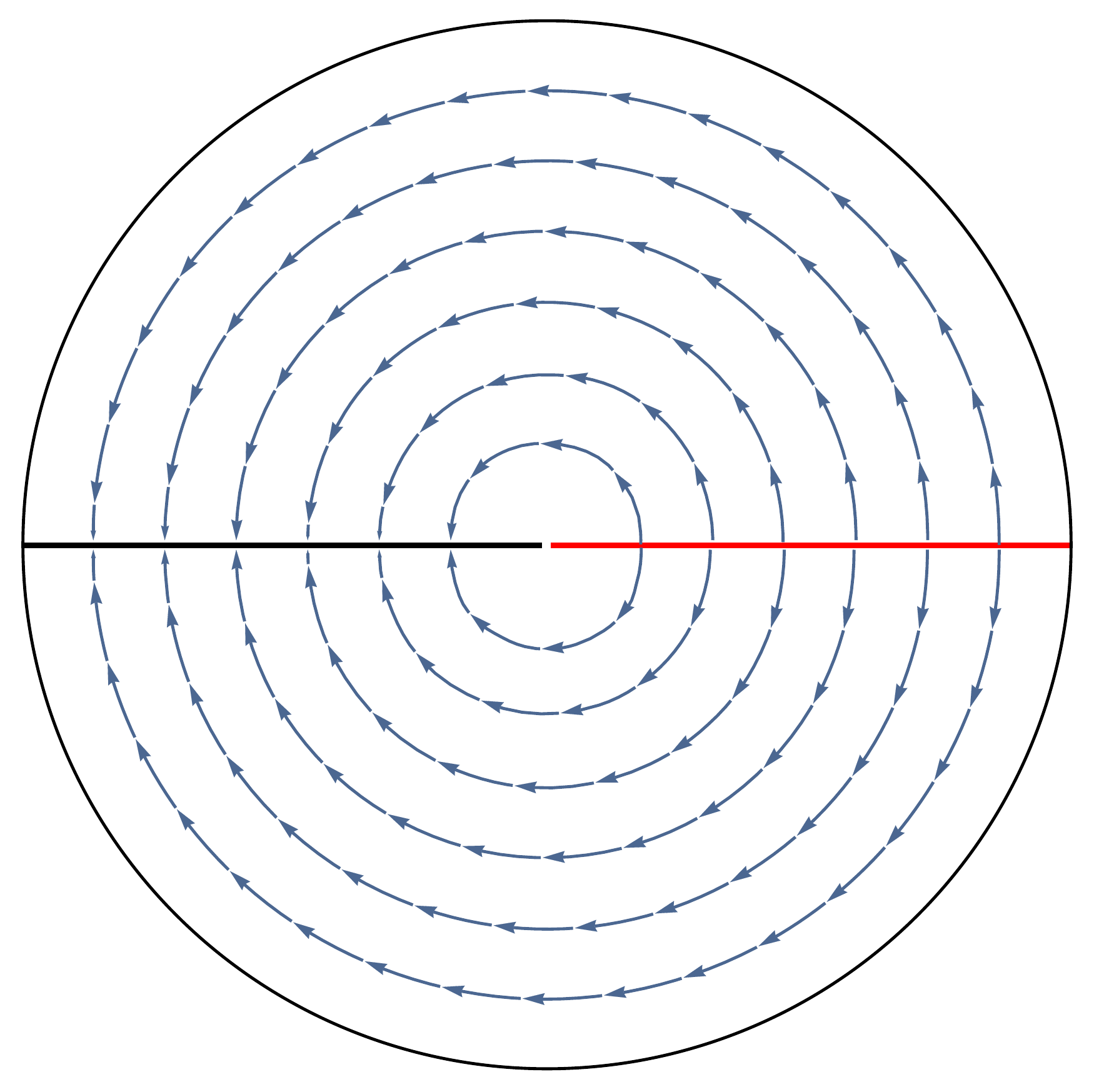}}
\hspace{1cm}\subfigure{
\includegraphics[scale=0.675]{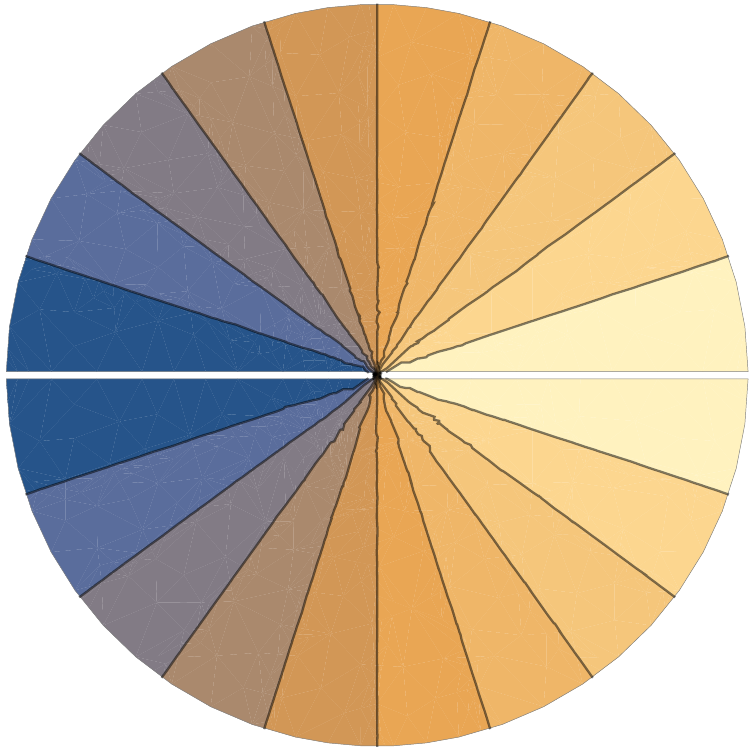}}
\caption[1]{\label{Fig_5}(Color online)
 Electric field lines and equipotentials for the ``monopole'' configuration (same as Fig.\,\ref{Fig_2}), but as calculated solving Laplace's equation numerically within a finite region of either square (upper pane) or circular (lower pane) boundary.}
\end{figure*}
%%%%%%%%%%%%%%%%%%%%%%%%%%%%%%%%%%%%%%%%%%%%%%%%%%%%%%%%%%%%%%%%%%%%%%%%%%%%%%%%
%

The potential is finally calculated using \texttt{NDSolve} and, from that, the electric field may be derived with the built-in function \texttt{Grad}. The results may be visualized with \texttt{ContourPlot} and \texttt{StreamPlot}. The usual options for graphics available in {\sc Mathematica} enable a convenient visual output, as shown in Fig.\,\ref{Fig_5}. We show electric field lines and equipotentials obtained for a realistic single slit ``monopole'' system over a finite board. 
On the one side, one may notice remarkable size effects for the square board. As expected, the idealized lines in Fig.\,\ref{Fig_2} deform so as to satisfy the above mentioned boundary conditions. Also as expected, one can verify that when solved in a circular domain, Laplace's equation {gives} a solution that is indistinguishable from the infinite idealized geometry. This {fact} could be anticipated in view of the uniqueness property of Laplace's equation. Thus, the boundary conditions for $\Phi$ on {each} half-circle (closed-region) are not different whether such region stands alone or is a part of the {upper/lower} half-plane. In physical terms, by cutting the paper so as to get the circle, one does not ``interrupt'' any ideal flux line and nothing changes within.

Just for completeness, we want to mention that the results of this section may be obtained in closed form by using conformal mapping \cite{ref:nehari} with relative ease (see supplementary material).

\subsection{Conjoined domains}
\label{sec_IIE}

\begin{figure*}[b]
\centering
\subfigure{
\raisebox{-2.ex}{\includegraphics[scale=.82]{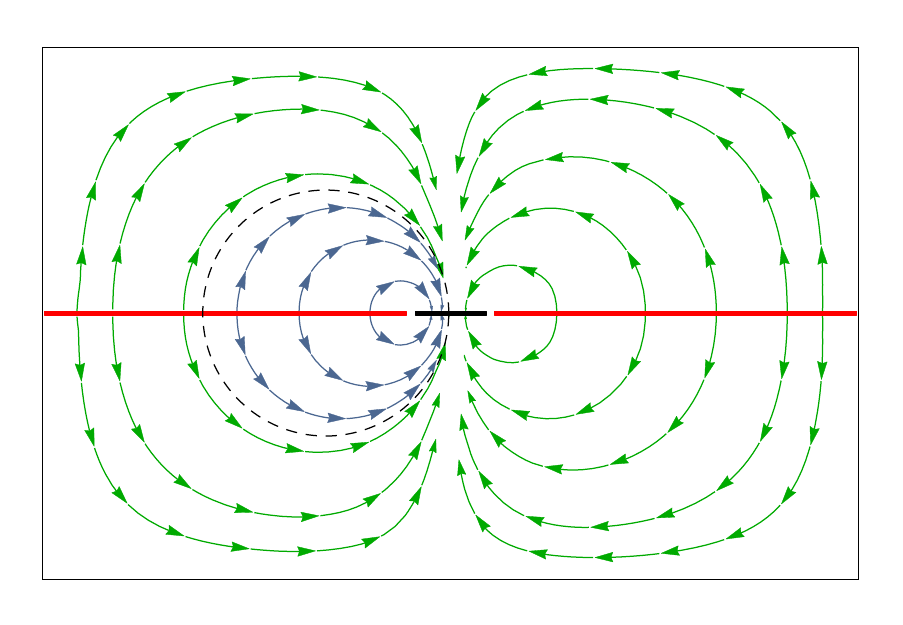}}}
\quad\subfigure{
\includegraphics[scale=0.36]{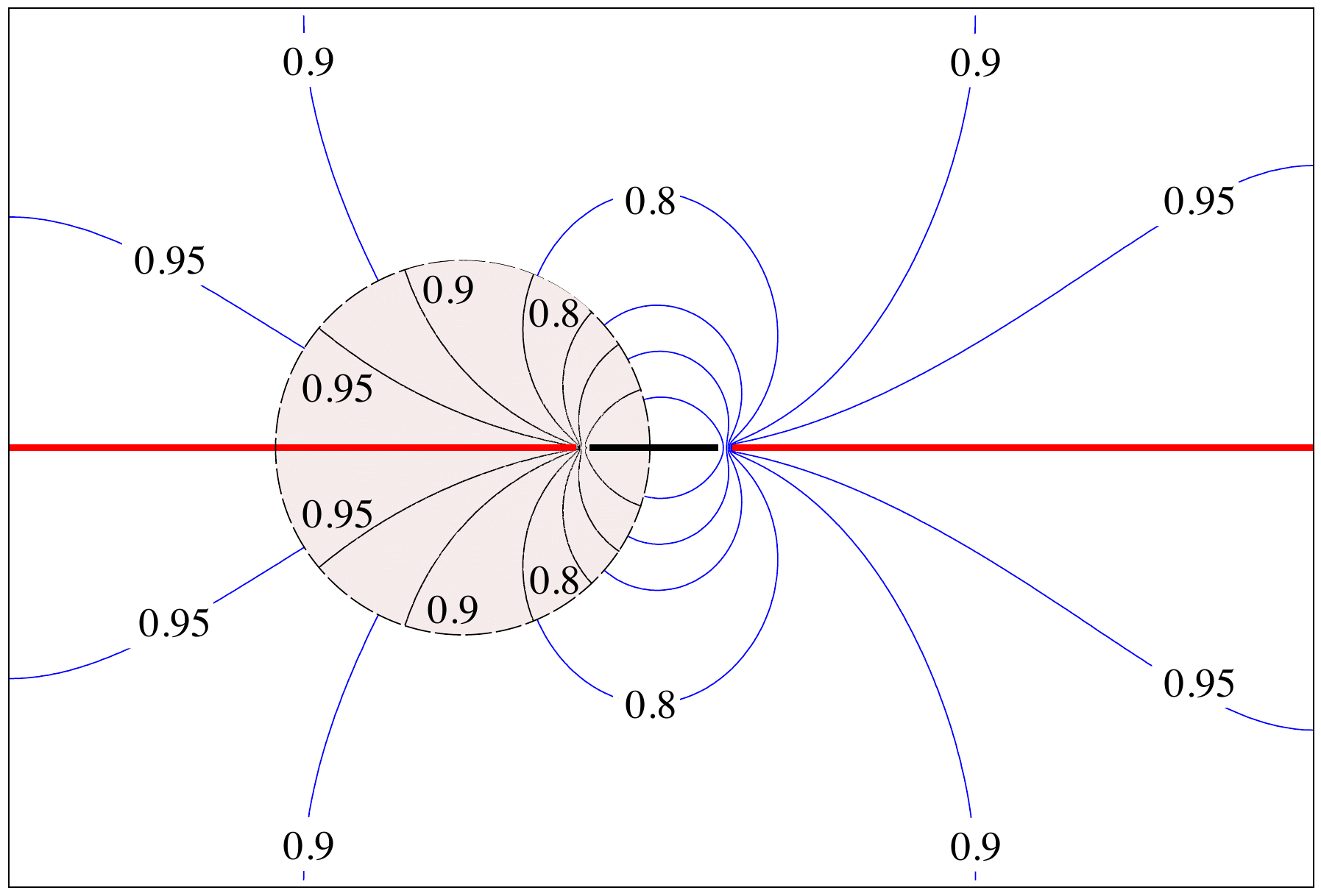}}
\caption{
\label{Fig_6}(Color online)
Theoretical electric field lines (left pane) and equipotentials (right) for the dipole configuration (same as Fig.\,\ref{Fig_3}) over a finite (rectangular) board. The board has been trimmed by cutting along the circumference indicated by  a dashed line. The electric potential within the circle is highlighted (but not zoomed!). Equipotentials (inner and outer) take the values $0.4\Phi_0, 0.6\Phi_0, 0.7\Phi_0, 0.8\Phi_0, 0.9\Phi_0, 0.95\Phi_0$. Some are labelled for convenience.}
\end{figure*}

Here, we show that one can go a step further if the idea of trimming  the paper is elaborated. Thus, taking advantage of the fact that in our ideal ``dipole'' configuration, electric field lines form circles around the slits (Eq.(\ref{eq:circle_elec})), we propose to trim the paper as shown in Fig.\,\ref{Fig_6} (avoiding to cut the silver strips!). By these means one may illustrate such relevant concepts in the electromagnetic theory as finite size, continuity, and electrical uncoupling. Also, the physical nature of the problem (stationary electrodynamics {\em vs.} electrostatics) will be clearly revealed.
%%%%%%%%%%%%%%%%%%%%%%%%%%%%%%%%%%%%%%%%%%%%%%%%%%%%%%%%%%%%%%%%%%%%%%%%%%%%%%%
%% FIGURE 5 
%%%%%%%%%%%%%%%%%%%%%%%%%%%%%%%%%%%%%%%%%%%%%%%%%%%%%%%%%%%%%%%%%%%%%%%%%%%%%%%
%%%%%%%%%%%%%%%%%%%%%%%%%%%%%%%%%%%%%%%%%%%%%%%%%%%%%%%%%%%%%%%%%%%%%%%%%%%%%%%%
%

Fig.\,\ref{Fig_6} displays the results of our numerical solution of Laplace's equation in a double domain, i.e.: square board with circular hole ($D_1$) and the complementary circular region ($D_2$). The boundary conditions for $\Phi (x,y)$ within each subdomain are again given by: (i) constant value on the conductors (Dirichlet) and (ii) null normal derivative on the boundaries (Neumann). Mathematically, one solves two problems, and obtains two solutions, the harmonic functions
\begin{eqnarray}
\Phi_1 (x,y) \, : \,  \{\Phi (x,y)\; | \; (x,y)  \in D_1\}
\nonumber\\
\Phi_2 (x,y) \, : \,  \{\Phi (x,y)\; | \; (x,y)  \in D_2\} 
\end{eqnarray}
Physically, this separation relates to the nature of the problem. Electric conduction is not allowed across the cut, and thus $D_1$ and $D_2$ are uncoupled regions (safe for sharing the values of the potential at the electrodes). Let us discuss the properties of the potential $\Phi (x,y)$ :
\begin{itemize}
\item[(i)] Within the detached circular region, one obtains the ``infinite geometry'' solution (as in {\em example 2}) \cite{ref:uniqueness}. See the supplementary material for more detail).
\item[(ii)] On the same board one visualizes the ``infinite geometry'' solution and the finite size related deformation. 
\item[(iii)]  The electric potential is notoriously discontinuous across the boundary between the conducting regions. At first, this property may be somehow striking. {Indeed, it} relates to the fact that the problem must be solved by parts. One cannot ``prolongate'' the solution $\Phi_1 (x,y)$ inwards. On the other hand, this creates no conflict from the physical point of view. $D_1$ and $D_2$ are different regions as concerns the physics of the problem (electric conduction), and the values of $\Phi(x,y)$ at nearby points are not constrained by continuity.
\end{itemize}

\section{Experiments in the laboratory}
\label{sec_III}

Below we present the practical counterpart of the systems described in the previous section. Relying on standard conductive paper supplies, silver paint and electrical instruments available in the Physics lab, we prepared a number of setups so as to assess the theoretical results. It will be shown that the accuracy of standard field-mapper kits \cite{ref:pasco} is sufficient to investigate the physical properties at a quantitative level.
These experiments may be readily implemented in the practical sessions for students in technical areas.

\subsection{Monopole configuration (single slit)}
\label{sec_IIIA}

\begin{figure*}[t]
\centering
\subfigure{\raisebox{6mm}{
\includegraphics[height=5.25cm]{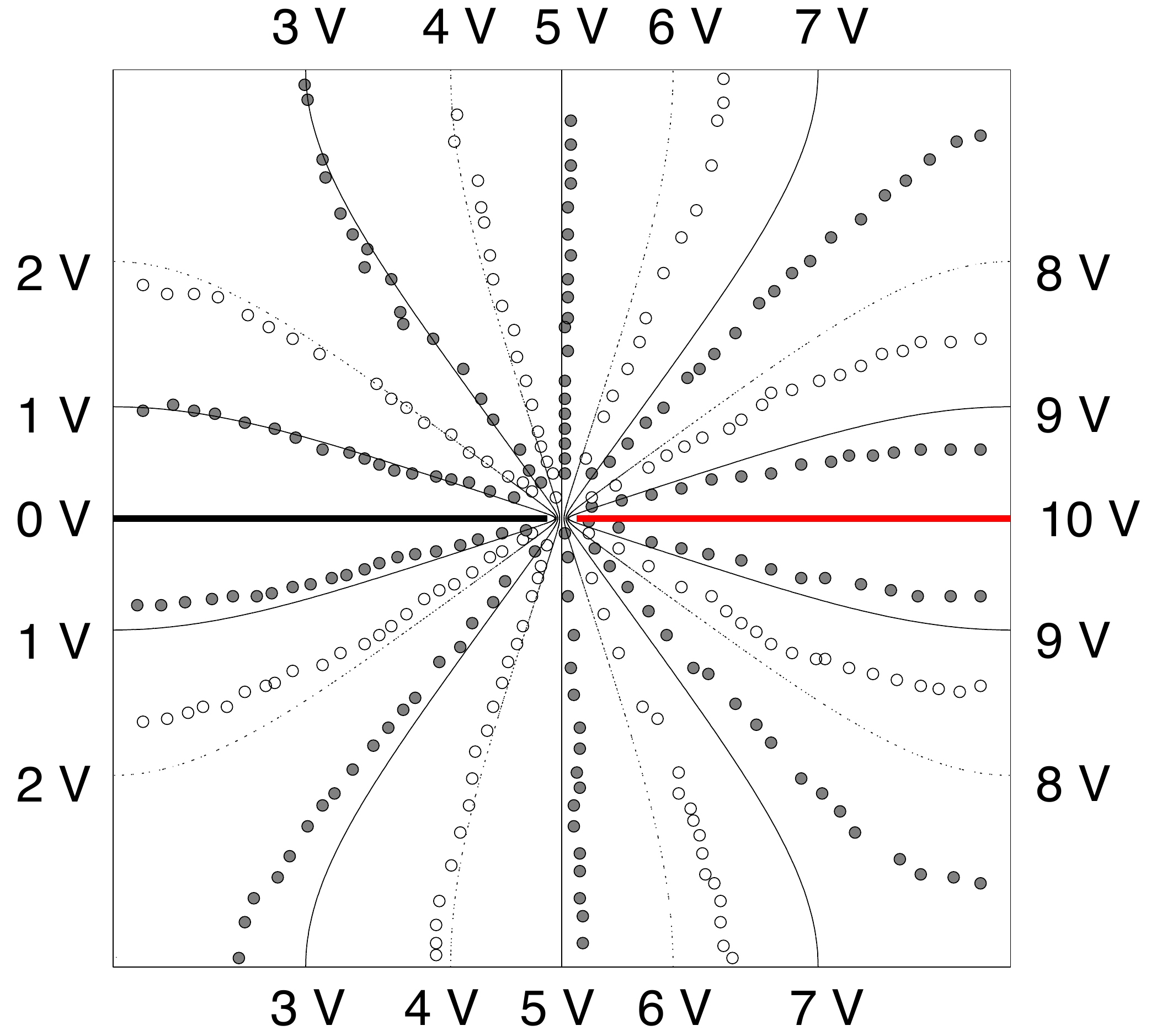}}}
%\noindent\rule{8cm}{0.4pt}
\quad\subfigure{
\includegraphics[height=6.5cm]{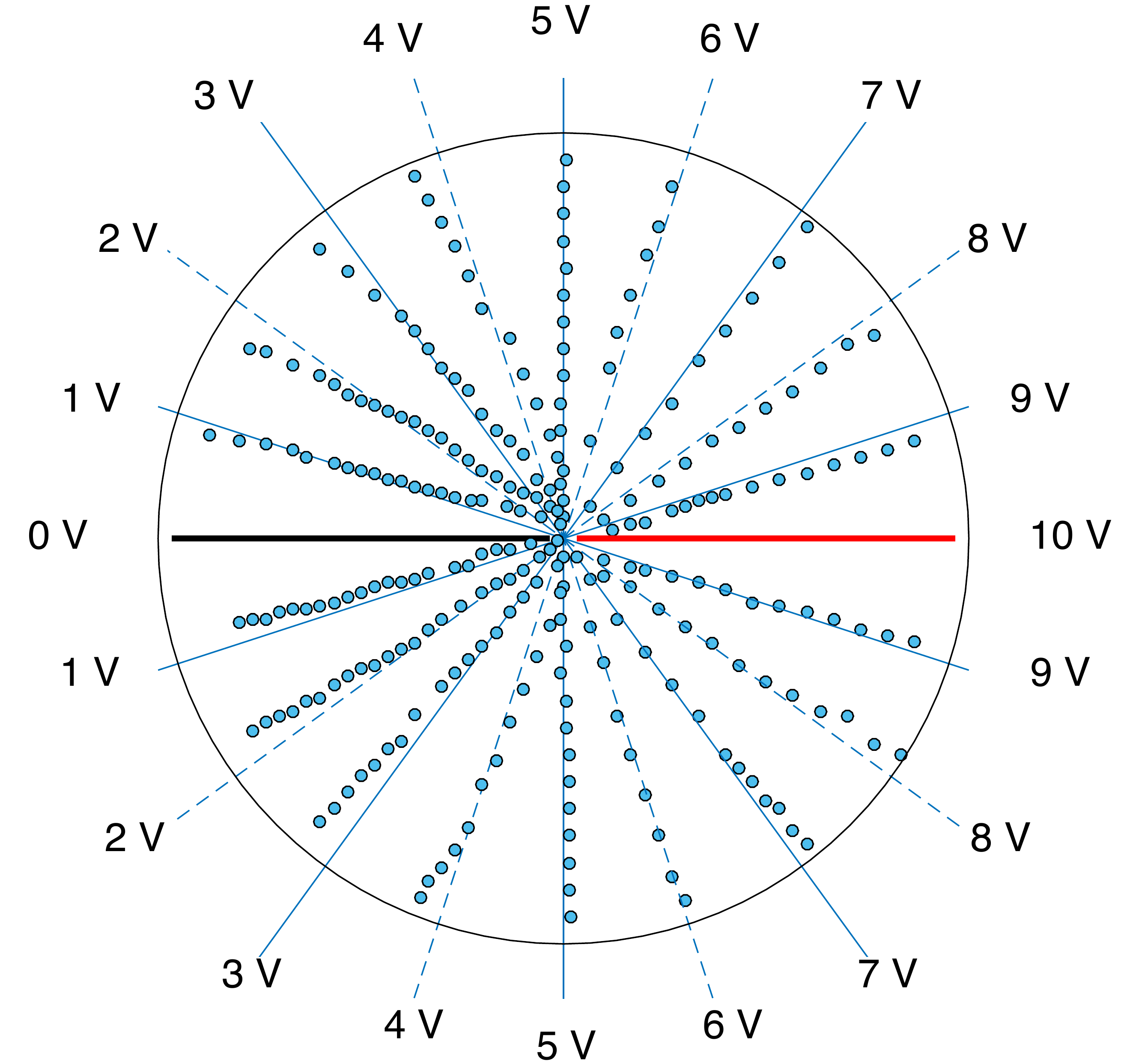}}
\caption{
\label{Fig_7}(Color online)
Experimental equipotentials obtained for a single slit configuration (monopole) with the conductors connected to a potential difference of 10 V. The left panel shows the results obtained for a a square shaped conducting paper (symbols). {Continuous lines correspond to the theoretical expectation.} To the right we display the modification of the map when the paper was cut and shaped as a circle.}
\end{figure*}

As deduced from the numerical analysis in Sec.\,\ref{sec_IID}, the ideal situation depicted in Fig.\,\ref{Fig_2} may be strongly deformed by influence of finite size effects. Such fact is notorious in Fig.\,\ref{Fig_5}. This situation was studied by using a $15\times 15\, cm^2$ board and silver painted electrodes of width $0.5\, cm$, with a central aperture of $0.5\, cm$. A potential difference of $10\, V$ was established between the electrodes by connecting to a DC power source. Equipotentials were recorded with help of a voltmeter and a pair of guiding rulers on top of the board. Results are shown in Fig.\,\ref{Fig_7}.
%
%%%%%%%%%%%%%%%%%%%%%%%%%%%%%%%%%%%%%%%%%%%%%%%%%%%%%%%%%%%%%%%%%%%%%%%%%%%%%%%
%% FIGURE 6 
%%%%%%%%%%%%%%%%%%%%%%%%%%%%%%%%%%%%%%%%%%%%%%%%%%%%%%%%%%%%%%%%%%%%%%%%%%%%%%%
%%%%%%%%%%%%%%%%%%%%%%%%%%%%%%%%%%%%%%%%%%%%%%%%%%%%%%%%%%%%%%%%%%%%%%%%%%%%%%%%
Together with the experimental points we plot the theoretical equipotential lines. Remarkably, by shaping the board as a circle, one checks the simple expression for infinite geometry

\begin{equation}
\label{eq:theor_V_mono}
\displaystyle{y = {\rm tan}\!\left(\frac{\pi \Phi}{10}\right) x}
\end{equation}

As one can observe in Fig.\,\ref{Fig_7}, the main features of our theoretical analysis are reproduced. For the case of the square board, equipotentials notoriously deviate from the linear behavior. As expected, deviation reflects the property of reaching the boundaries at right angle. Concerning the quantitative validity, one can observe that theoretical equipotentials are nicely followed by the experimental points, especially for the case of the circular board. From our view, deviations can be assigned to the limitations in the experimental precision, imperfect homogeneity of the conductive paper, etc.

{\subsection{Dipole configuration (double slit)}}
\label{sec_IIIB}
As foreseen in our simulations (Sec.\,\ref{sec_IIE}) one may expect that ideal dipole lines will be revealed in a real double slit system by trimming the paper in some precise manner. Specifically, by cutting the paper along a circumference given by Eq.(\ref{eq:circle_elec}) equipotentials should follow the ideal circular shape of Eq.(\ref{eq:circle_pot}) {within the circle}. Also, they should change discontinuously across the cut, taking very different values at nearby points.
%
%%%%%%%%%%%%%%%%%%%%%%%%%%%%%%%%%%%%%%%%%%%%%%%%%%%%%%%%%%%%%%%%%%%%%%%%%%%%%%%
%% FIGURE 7 
%%%%%%%%%%%%%%%%%%%%%%%%%%%%%%%%%%%%%%%%%%%%%%%%%%%%%%%%%%%%%%%%%%%%%%%%%%%%%%%
\begin{figure*}[t]
\centering
\includegraphics[scale=0.38]{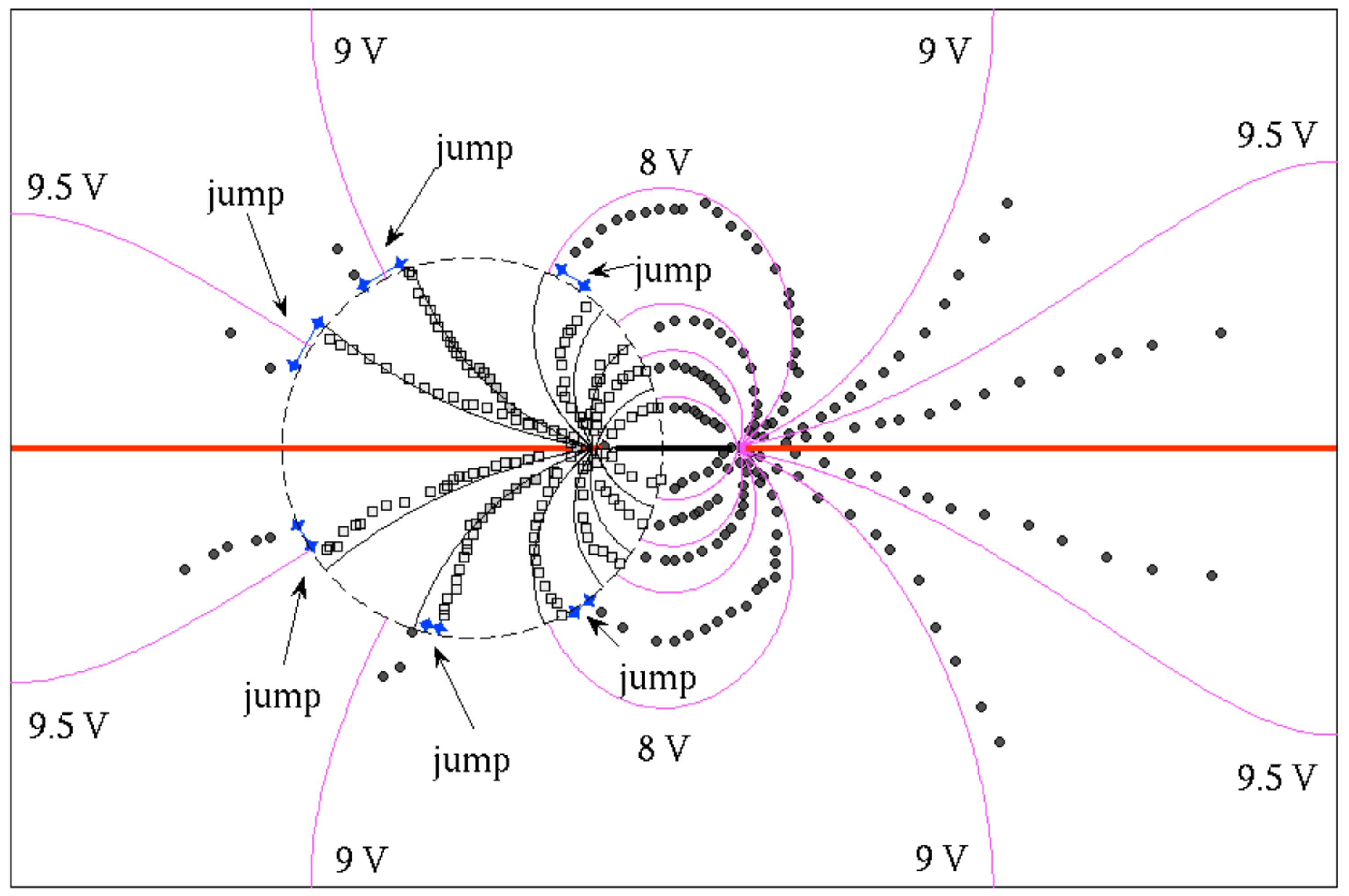}
\caption{
\label{Fig_8}(Color online)
Experimental equipotentials (symbols) for a double slit configuration with the conductors connected to a potential difference of 10 V. The conducting paper was trimmed along the dashed circumference. Continuous lines correspond to the theory. Discontinuities in the measured potential are highlighted. For clarity, some lines are not labelled ($4 V, 6 V, 7 V$).}
\end{figure*}
%%%%%%%%%%%%%%%%%%%%%%%%%%%%%%%%%%%%%%%%%%%%%%%%%%%%%%%%%%%%%%%%%%%%%%%%%%%%%%%%
%
All the above facts could be clearly observed in the dedicated field mapping setup shown in Fig.\,\ref{Fig_8}. Here, we used a sheet of size $42\times 30\, cm^2$ with a central silver strip of width $0.5\, cm$  interrupted by apertures of $0.5\, cm$ positioned at $(x=\pm 2.5\, ,\, y = 0\,)\, cm$. The central conductor was connected to the negative of the battery, and the lateral strips to the positive ($10\, V$). By means of a scalpel, a circumference of radius $6.6\, cm$ and center at $(x=-7\, ,\, y = 0\,)\, cm$ was cut, avoiding to touch the silver conductors.

We notice that observations reasonably correspond to the theoretical expectations, mainly within the circular region, in which the experiment practically reproduces the theoretical curves.

\bigskip
\section{Discussion}
\label{sec_IV}

We propose an alternative to the widespread field-mapping experiments in introductory and intermediate courses of electricity. It is shown that, by drawing the electrodes as long lines interrupted by tiny slits and shaping the board in precise manner, a number of physical properties may be observed and quantified: complementarity between electric field lines and equipotentials, finite-size effects, continuity/discontinuity of the potential, and electrical coupling/uncoupling between adjacent regions. Some interesting topics for testing the students' level of understanding would be: (i) is it correct to obtain radial equipotentials in the single slit configuration? (ii) why is the role of electric field lines and equipotentials interchanged? (iii) why is the response so different when one changes the shape of the underlying board? (iv) for the double slit system: is it correct to obtain a discontinuous potential? (v) why does this occur when we cut the conducting paper?

When addressed to the intermediate undergrad courses, these experiments pave the path for acquaintance with complex variable methods in electromagnetism. Thus, starting with the single slit configuration, one may straightforwardly (just by inspection) obtain the complex potential $\psi_m (z)=\zeta_m (z)+ i\Phi_m (z)$ for the system and later use it to describe higher elements by superposition.
As an immediate reward, one gets simple analytic expressions for the electric field lines, i.e.: $\zeta (x,y) = constant$, straightforwardly  derived upon knowing $\Phi (x,y)$ because these functions are harmonic conjugates (consider f.i. $\Phi_d (x,y)$ and $\zeta_d (x,y)$ in Eq.(\ref{eq:complex_dipo})). The availability of handy algebraic expressions for the function $\zeta (x,y)$ has allowed to envisage experimental patterns that either avoid or emphasize the effect of boundary conditions. Thus, if one trims the conducting paper along a line given by $\zeta (x,y) = constant$, the inner domain exactly reproduces the infinite geometry solution. Remarkably, this can be realized in practice with a standard laboratory kit. 

We have also worked out numerical calculations by means of the popular software {\sc Mathematica} \cite{ref:mathematica}. A fairly simple code \cite{ref:code} allows to solve Laplace's equation, double-check closed form solutions, and anticipate their form in the more complicated cases.

%%%%%%%%%%%%%%%%%%%%%%%%%%%%%%%%%%%%%%%%%%%%%%%%%%%%%%%%%%%%%%%%%%%%%%%%%%%%%%%
%% FIGURE 8 
%%%%%%%%%%%%%%%%%%%%%%%%%%%%%%%%%%%%%%%%%%%%%%%%%%%%%%%%%%%%%%%%%%%%%%%%%%%%%%%
\begin{figure*}[t]
\centering
\includegraphics[scale=0.4]{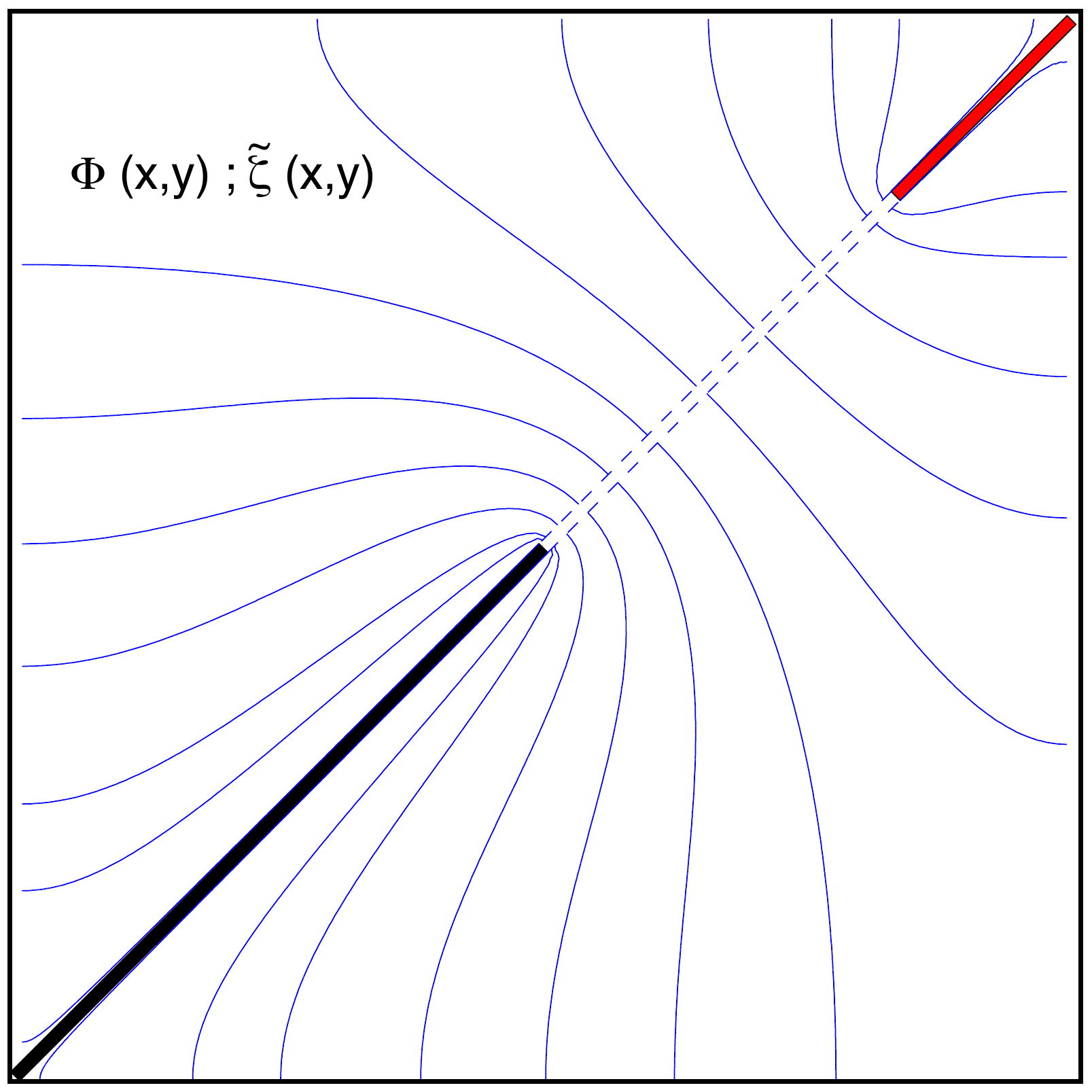}
\caption{
\label{Fig_9}(Color online)
Representation of either (i) the equipotentials ($\Phi = constant$) in a system with two aligned electrodes with finite aperture or (ii) the electric field lines ($\tilde{\zeta} = constant$) around the dotted conductor surrounded by a constant potential square frame.}
\end{figure*}
%%%%%%%%%%%%%%%%%%%%%%%%%%%%%%%%%%%%%%%%%%%%%%%%%%%%%%%%%%%%%%%%%%%%%%%%%%%%%%%%

Further work is in progress in the following terms: is there a method that generalizes our examples of dual systems with interchanged roles between the electric field lines and equipotentials? In particular: is the duality restricted to systems with tiny slits?
The basic facts may be examined with help of Fig.\,\ref{Fig_9}. Straightforwardly, one can recognize the equipotentials on a square board with two inclined (and aligned) electrodes, separated by a finite gap. Recall that equipotentials are perpendicular to the edges of the board, and also to the dashed line along the gap ($\partial_{n}\Phi (x,y)=0$), whereas one has constant $\Phi$ on the electrodes. On the other hand, the same contour plot would be obtained for the electric flux function $\tilde{\zeta} (x,y)$ over the region that surrounds a central electrode defined by the dashed rectangle, framed by a peripheral conducting square. The key point is that $\Phi (x,y)$ and $\tilde{\zeta}(x,y)$ are harmonic functions over the same domain and satisfy the same boundary conditions \cite{ref:uniqueness}.
In practice, notice that the electrodes of the $\Phi-$system are defined along electric field lines of the $\tilde{\zeta}-$system, and that, in general, one would need to cut the paper along the dashed line (central conductor) so as to ensure the Neumann condition for $\Phi$, which in Fig.\,\ref{Fig_9} follows by symmetry.

\ack

M. L. Ram\'on is thankfully acknowledged for her skill and patience in the arrangement and realization of the experiments. Partial financial support from the Spanish MINECO under project ENE2017-83669-C4-1-R is acknowledged.


\begin{thebibliography}{99}
\bibitem{ref: multivalued} {For the sake of rigour, and considering that multivalued functions will be involved, one should have some care with the definition of $\psi$. The necessary considerations will be done for each specific case. A general mathematical framework would involve the consideration of branch cuts, which is beyond the scope of this paper}
\bibitem{ref:code} {The {\sc Mathematica} notebook is available upon request to the authors}
\bibitem{ref:uniqueness} {Mathematically, this is a consequence of the uniqueness of solutions to the Laplace equation in a bounded domain with mixed boundary conditions. In the current situation, one has Dirichlet conditions on the conductor, and Neumann conditions along the cut}
\end{thebibliography}
\end{document}